\documentclass[aps, pra, twocolumn, amssymb, showpacs]{revtex4-1}
\usepackage{ulem}
\usepackage{graphicx}
\usepackage{bm}
\usepackage{hyperref}
\usepackage{float}
\usepackage{xcolor}
\usepackage{mathtools}
\newcommand{\dif}{\mathrm{d}\;\!\!}

\newcommand {\Seq}	{Schr\"odinger equation}
\newcommand {\Schr}	{Schr\"odinger}
\newcommand {\Wn} 	{Wannier}
\newcommand {\BZ} 		{Brillouin zone}

\newcommand {\Mt}	{Mathieu}
\newcommand {\WW}	{Weisskopf-Wigner}
\newcommand {\FC}	{Franck–Condon}
\newcommand {\Bl}	{Bloch}

\newcommand{\Wk}	{Wick}

\newcommand{\ket}[1]{\left|#1\right\rangle}
\newcommand{\bra}[1]{\left\langle#1\right|}
\newcommand{\esc}[2]{\left\langle#1\left|\right.#2\right\rangle}

\newcommand{\appropto}{\mathrel{\vcenter{
  \offinterlineskip\halign{\hfil$##$\cr
    \propto\cr\noalign{\kern1pt}\sim\cr\noalign{\kern-2pt}}}}}

\newcommand{\FA}{\mathcal{F}\hspace{-2pt}A_{\hspace{1pt}q}}

\newcommand{\DispersionRelation}{\varepsilon_q}
\newcommand{\Energy}{E}
\newcommand{\DispersionRelationNoIndices}{\varepsilon}
\newcommand{\Integral}{I}
\newcommand{\EmitterResidue}{\alpha}
\newcommand{\PolaritonResidue}{r}
\newcommand{\Transfer}{T}

\begin{document}
\title{Multiband and array effects in matter-wave-based waveguide QED}
\author{Alfonso Lanuza}
\thanks{alfonso.lanuza@stonybrook.edu}
\author{Joonhyuk Kwon}
\author{Youngshin Kim}
\author{Dominik Schneble}
\affiliation{Department of Physics and Astronomy, Stony Brook University, Stony Brook, NY 11794-3800, USA}

\date{\today}

\begin{abstract}
Recent experiments on spontaneous emission of atomic matter waves \cite{Krinner2018,Stewart2020} open a new window into the behavior of quantum emitters coupled to a waveguide. 
Here we develop an approach based on infinite products to study this system theoretically, without the need to approximate the band dispersion relation of the waveguide. We solve the system for a one-dimensional array of one, multiple and an infinite number of quantum emitters and compare with the experiments. This leads to a detailed characterization of the decay spectrum, with a family of in-gap bound states, new mechanisms for enhanced Markovian emission different from superradiance, and the emergence of matter-wave polaritons.
\end{abstract}

\maketitle

\section{Introduction}
\label{sec:intro}
Experimenting with the light-matter interface at the quantum level \cite{Hood2016,Dibyendu2017,Carusotto2020} has led to the discovery of new exciting phenomena, including implementations of photonic quantum matter \cite{Chang2018,Carusotto2013},
 effective long-range interactions between neutral atoms \cite{Douglas2015}, chiral quantum optics \cite{Mitsch2014}, nondestructive photon counting \cite{Malz2020} and quantum logic gates \cite{Zheng2013,Paulisch2016} among others. The plethora of these developments contrasts with the conceptual simplicity of coupling one or many quantum emitters to a photonic waveguide (a.k.a.~waveguide QED); photons are the ideal carriers of quantum information as they can travel large distances without colliding with one another, while atoms can access and store this information by absorbing the photons.\\

Implementations using ultracold atoms in optical lattices, originally proposed by \cite{deVega2008}, challenge this paradigm as they reproduce the same physics while switching the roles of matter and light which now become the radiation and emitters, respectively. This comes with advantages, such as accessibility to different parameter regimes and a high tunability of the system \cite{Navarrete2011, Stewart2017} with the potential to enable state-of-the-art applications including the simulation of giant atoms \cite{GTudela2019}, perfect subradiance \cite{GTudela2018B} and topological effects \cite{Bello2019}. 
The experimental realization of these systems has been achieved recently \cite{Krinner2018, Stewart2020} resulting in the observation of phenomena such as tunable Lamb shifts, bound-state beats, non-Markovian decay dynamics and time-of-flight pictures of the bound states.\\

In the analysis of waveguide-QED systems \cite{Rzazewski1982, Lewenstein1988, Kofman1994, Lambropoulos2000,Berman2010}, a key idea is that the emission properties can be drastically altered by manipulating the mode distribution of the radiation \cite{Bykov1975, Yablonovich1987, John1987}. However, due to the difficulty of a multiband analysis \cite{Bykov1975}, the emission of matter wave radiation has only been analyzed for the cases of a single emitter coupled to modes with one parabolic band edge \cite{deVega2008,Stewart2017,Krinner2018} or one sinusoidal band \cite{Calajo2016,Stewart2020}. In this paper, we go beyond these restrictions and present how to evaluate the exact lattice functions required to solve the dynamics. Using techniques of complex analysis, we find that this can be done in an efficient way by using infinite products. These allow us to solve the system with broader generality (i) by considering the action of multiple emitters in the system; (ii) by accounting for the full band dispersion relation for the radiated modes; (iii) going beyond the Markovian regime, as we solve for arbitrarily high couplings; and (iv) also accounting for the finite size of the emitters. For simplicity, and in agreement with the conditions of the experiments \cite{Krinner2018,Stewart2020}, we restrict ourselves to 1D non-interacting systems within the single excitation subspace.\\

This paper is organized as follows. In Sec.~\ref{sec:systemHamiltonian}, we give an introduction to the ultracold atom platform and its connection with the Hamiltonian for spontaneous emission of matter waves. In Sec.~\ref{sec:dynamics}, we derive the formal solution of the equations of motion for multiple emitters. These equations can be evaluated using the infinite-product representations presented in Sec.~\ref{sec:complexEplane}. We focus on the case of a single emitter in Sec.~\ref{sec:N=1}, where we describe the system spectrum in detail and compare the analysis results with the experiments. In Sec.~\ref{sec:BoundStates} we study the form of the bound states, consisting of matter waves dynamically anchored to the emitter. Finally, in Sec.~\ref{sec:polaritons} we review the case of an infinite array of emitters and the emergence of polaritons that result from the hybridization between the emitter array and the potential in which the matter waves propagate.\\

\section{The System Hamiltonian}\label{sec:systemHamiltonian}

We consider a two-level  atom (states ``red" $\ket{a}$  and ``blue" $\ket{b}$) of mass $m$ in a 1D state-dependent optical lattice of recoil momentum $k$ and potential $\mathbf{V}_{a(b)}(z)=V_{a(b)}\sin^2 (kz)$, as experimentally realized in \cite{Krinner2018, Stewart2020}. When the atom is in $\ket{a}$, the lattice potential is deep ($V_a\gg E_r=(\hbar k)^2/2m$) and the atom is locally confined to the harmonic-oscillator ground state $\phi_j(z)$ in site $j$, with negligible tunneling between sites. An oscillating external field of frequency $\omega_\mu$ couples the atom to $\ket{b}$, which experiences a much shallower  potential $\lvert V_b \rvert\ll V_a$, with tunneling $J_b/\hbar$.\\

\begin{figure}[t]
	\begin{center}
	\includegraphics[width=1\columnwidth]{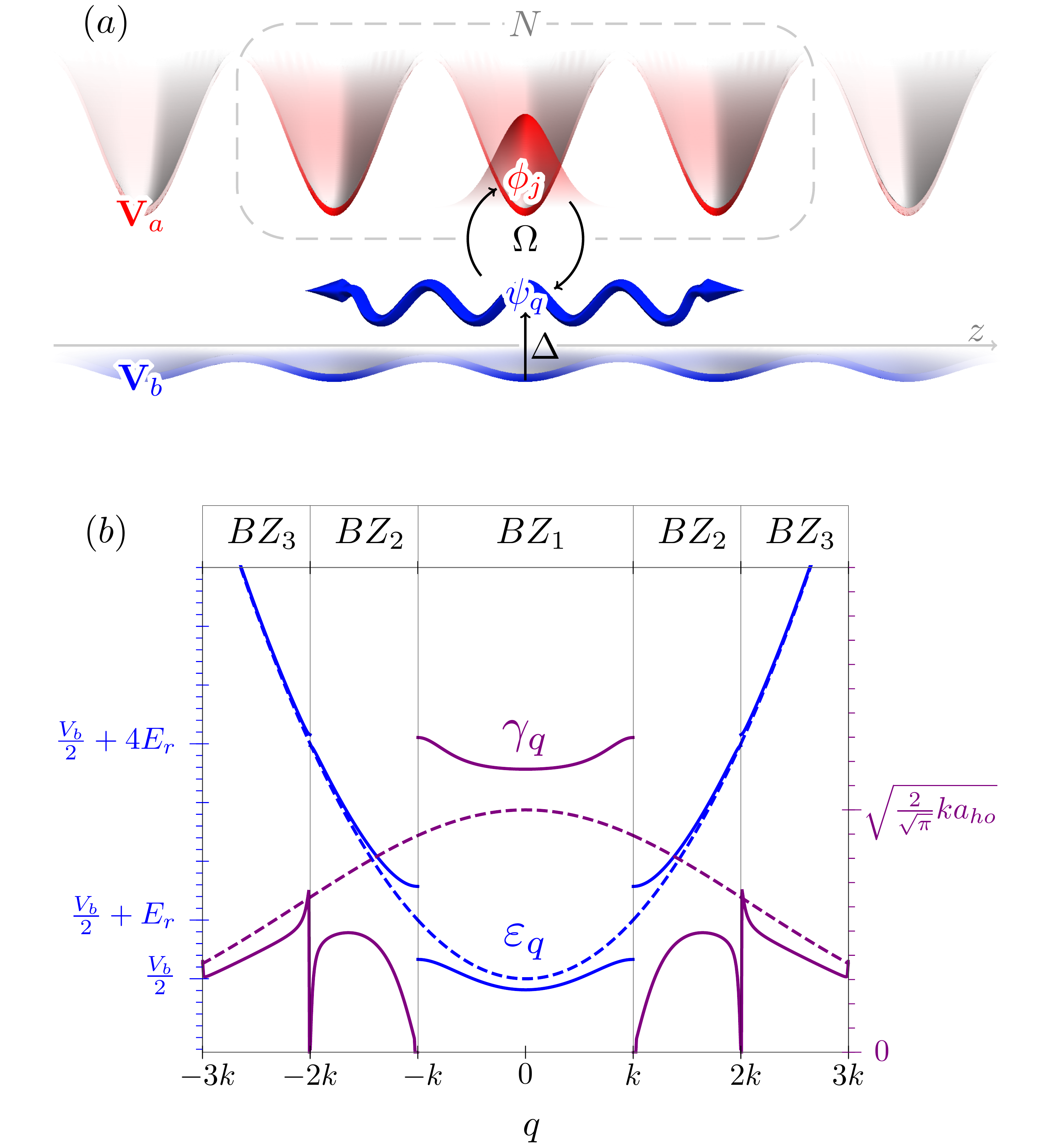}
	\caption{Representation of the system. (a) The system in position space. Gaussian wavefunctions $\phi_j$ representing the ground states of altogether $N$ considered wells of a deep lattice $\mathbf{V}_a(z)$ are coupled to Bloch waves $\psi_q$ of a shallower lattice $\mathbf{V}_b(z)$, through a coupling of strength $\Omega$ and detuning $\Delta$ from  $\mathbf{V}_b(0)$. (b) Energy bands $\DispersionRelation$ (in blue) and \FC\ overlap $\gamma_q$ (in purple) in the extended zone scheme. Solid lines are for the case $V_b=2.5 E_r$, and the dashed lines for the free-particle case $V_b=0$.}
	\label{fig:System}
	\end{center}
\end{figure}

The motional states $\ket{1^b_{q}}$ of the $\ket{b}$ atoms are the Bloch waves $\psi_{q}(z)$ with lattice momentum $q\in(-\infty,+\infty)$ which are solutions of the \Mt\ equation \cite{MathieuBook}. We choose to normalize these to a single \BZ\ ($BZ$) but work in the extended zone scheme for a more direct connection with the free-particle case. We take their total energy $\hbar\omega_{b,q}$ as the sum of the  band energy $\DispersionRelation$ (see Fig.~\ref{fig:System}(b)) of  $\mathbf{V}_{b}(z)$ and the internal energy $\hbar\omega_{b}$. We note that for diverging energy, the dispersion relation $\DispersionRelation$ approximates that of a free particle subject to the constant average potential $V_b/2$; this will become important in Section \ref{sec:complexEplane}.\\

If the atom is initially localized in a single well, during short evolution times $t\ll \hbar/J_b$ the system is a simulator for spontaneous emission of a photon by an isolated quantum emitter in a photonic crystal \cite{deVega2008, Stewart2017}. However, for longer times, the $\ket{b}$ atom propagates sufficiently far to be reabsorbed by neighboring lattice sites of $\mathbf{V}_{a}$, such that dynamical array effects become noticeable, as was experimentally observed in \cite{Krinner2018, Stewart2020}. These array effects are detrimental to the Markovianity of the system \cite{Cascio2019} and they are especially acute in the ultracold platform due to the strong retardation between emitters.
For such a system of $N$ lattice sites the uncoupled Hamiltonian is given by
\begin{equation}\label{eq:H0}
\hat H_0=\sum_{j=1}^N \hbar\omega_a \hat{a}_j^\dagger\hat{a}_j+\sum_q  \hbar\omega_{b,q} \hat{b}_q^\dagger \hat{b}_q,
\end{equation}
where $\hat{a}_j^\dagger=|1^a_{j}\rangle\bra{0}$ creates an $\ket{a}$ atom in the $j$-th site from the vacuum $\ket{{0}}$, $\hat{b}_{q}^\dagger=|{1^b_{q}}\rangle\bra{0}$ creates a $\ket{b}$ atom with lattice momentum $q$, and the explicit sum in modes is $\sum_q\equiv\int_{-\infty}^{+\infty}\frac{\dif q}{2k}$ for the chosen normalization and zone scheme. The coupling part of the Hamiltonian (in the interaction picture) takes the form
\begin{equation}\label{eq:Hab}
\hat{H}_{ab}=\sum_{j=1}^N\sum_q \frac{\hbar\Omega}{2}e^{i(\DispersionRelation/\hbar-\Delta )t-iqz_j}\gamma_{q} \hat{a}_j\hat{b}^\dagger_{q}+\text{H.c.},
\end{equation}
where $\Omega$ is the coupling between the two internal states, $z_j = j\pi/k$ is the position of the $j$-th site, and $\gamma_q=\esc{\psi_q}{\phi_0}$ (see Fig.~\ref{fig:System}(b)) is the \FC\ overlap between the tightly confined wavefunction and the Bloch wave at $q$, and $\Delta=\omega_\mu+\omega_a-\omega_b$ is the detuning of the coupling field \cite{Note1}.  The Hamiltonian thus takes the form of a system of \WW~Hamiltonians \cite{Weisskopf1930}, in which $\hbar\Delta$ plays the role of the excitation energy of a quantum emitter.\\


\section{Decay of an $N$-emitter array}
\label{sec:dynamics}

Generally, the state of an atom in the state-dependent lattice can be expressed as a linear combination of \Wn\ states (harmonic-oscillator ground states) and \Bl\ waves
\begin{equation}\label{eq:psi}
\ket{\psi(t)}=\sum_{j=1}^N A_j(t)\ket{1^a_{j}}+\sum_q  B_q(t)\ket{1^b_{q}}.
\end{equation}
The time evolution in this picture, $i\hbar {\partial_t}|\psi (t)\rangle ={\hat {H}}_{ab}|\psi (t)\rangle$, is given by
\begin{equation}\label{eq:EOMsystem}
\left\lbrace\begin{array}{l}
\dot A_j(t)=-\sum_q \frac{i\Omega}{2}e^{iqz_j-i(\DispersionRelation/\hbar-\Delta) t}\gamma_q^* B_q(t)\\
\dot B_q(t)=-\sum_{j} \frac{i\Omega}{2}e^{-iqz_j+i(\DispersionRelation/\hbar-\Delta) t}\gamma_q A_j(t).\\
\end{array}\right.
\end{equation}
Following a standard approach in the field \cite{Kofman1994, Lambropoulos2000,deVega2008,Navarrete2011,Quang1997,Stewart2017} and extending it to a system of sites, the system of differential equations can be solved via Laplace transform $\tilde A(s)\equiv \mathcal{L}\lbrace A(t)\rbrace$. Considering processes of spontaneous emission ($B_q(t=0)=0$), Eqns.~(\ref{eq:EOMsystem}) transform into
\begin{equation}
\left\lbrace\begin{array}{l}
s\tilde A_j(s)-A_j(0)=-\sum_q \frac{i\Omega}{2}e^{iqz_j}\gamma_q^* \tilde B_q\left(s+i(\DispersionRelation/\hbar-\Delta)\right)\\
s\tilde B_q(s)=-\sum_j \frac{i\Omega}{2}e^{-iqz_j}\gamma_q \tilde A_j\left(s-i(\DispersionRelation/\hbar-\Delta)\right).\\
\end{array}\right.
\end{equation}
Substituting the second equation into the first yields
\begin{equation}\label{eq:Atildes}
s\tilde A_j(s)-A_j(0)=-\sum_J \tilde G_{j\text{-}J}(i\hbar s+\hbar\Delta)\:\tilde A_J(s),
\end{equation} with the bath correlation function
\begin{equation}\label{eq:defGtilde}
\tilde G_{j\text{-}J}(\Energy)\coloneqq \frac{\hbar\Omega^2}{4i}\sum_q\lvert\gamma_q\rvert^2\frac{e^{iq(z_j-z_J)}}{\DispersionRelation-\Energy}.
\end{equation}

To proceed, it is convenient to express the Bloch waves $\psi_q$ entering the calculation of the \FC~factor $\gamma_q$ in the basis of plane waves
\begin{equation}
    \label{eq:upsilons}
    \psi_{q(\Energy)}(z)=\psi_{q(\Energy)}(0) \sum_{n=-\infty}^{+\infty}\upsilon_n(\Energy)e^{i(q(\Energy)+2nk)z},
\end{equation}
and to replace the sum over lattice momenta $q$ with an integration over the associated energy $E$. After integration in the complex plane (see Appendix~\ref{app:planeDecomposition}), this yields
\begin{widetext}
\begin{equation}\label{eq:modalGtilde}
\begin{aligned}
\tilde G_{j\text{-}J}&(\Energy)=
\frac{\hbar\Omega^2 a_{ho}\pi^{3/2}}{8k}
\rho(\Energy)\psi_{q(\Energy)}^2(0)\sum_{m,r=-\infty}^{+\infty}\upsilon_m(\Energy)\upsilon_r(\Energy)
e^{-{(q(\Energy)+k(m+r))^2a_{ho}^2}}
\\
&\:\times \left[e^{-iq(\Energy)(z_J-z_j)}\operatorname{erfc}\left(-i{(q(\Energy)+2kr)a_{ho}}+\frac{z_J-z_j}{2a_{ho}}\right)
+(z_J\leftrightarrow z_j)\right],
\end{aligned}
\end{equation}
\end{widetext}
where lattice momentum $q(\Energy)$ is the positive branch of the inverse function of the analytic extension of $\DispersionRelation$, $a_{ho}$ is the harmonic oscillator length (i.e. $\phi_0(z)=\exp[-z^2/(2 a_{ho}^2)]/\sqrt[4]{\pi a_{ho}^2}$), \hypertarget{(DoS)}{${\rho(\Energy)=2\dif q(\Energy)/\dif\Energy}$} is the density of Bloch states (DoS), with the prefactor of 2 accounting for the existence of left and right-movers, and $\operatorname{erfc}(x)=2/\sqrt{\pi}\int_x^\infty \exp(-y^2){\dif y}$ is the complementary error function.\\

We note that, because of the Gaussian envelope $\exp\left[{-{(q(\Energy)+k(m+r))^2a_{ho}^2}}\right]$ the infinite sum in Eqn.~\eqref{eq:modalGtilde} converges rapidly, and for a tightly confining emitter  ($\lvert a_{ho}q(\Energy)\rvert,\:a_{ho}k\ll1$), the expression above simplifies greatly. It  follows from (\ref{eq:upsilons}) that $\sum_m \upsilon_m(\Energy)=1$, and the $\operatorname{erfc}(x)$ in the formula approximates a step function. Hence,
\begin{equation}\label{eq:tightGtilde}
\tilde G_{j\text{-}J}(\Energy)\approx
\frac{\hbar\Omega^2}{4}\frac{a_{ho}\pi^{3/2}}{k}
\rho(\Energy)\psi^2_{q(\Energy)}(0)
 e^{i q(\Energy)\lvert z_J-z_j\rvert}.
\end{equation}\\

The linear system of equations~(\ref{eq:Atildes}) can be solved for finite $N$ with basic linear algebra (the infinite case $N=\infty$ is treated in detail in Sec.~\ref{sec:polaritons}), with the solution in matrix form written as $\tilde{\mathbf{A}}(s)=[s\mathbf{I}+\tilde{\mathbf{G}}(i\hbar s+\hbar\Delta)]^{-1}\mathbf{A}(0)$. Using the \Wk-rotated variable $\Energy=i\hbar s+\hbar\Delta$ in the inverse Laplace transform then yields
\begin{equation}\label{eq:inverseLaplace}
\begin{aligned}
    \mathbf{A}(t)=\frac{-1}{2\pi i}\int_{-\infty+i 0^+}^{+\infty+i 0^+}
   & \left[(\Energy-\hbar\Delta)\mathbf{I}+i\hbar\tilde{\mathbf{G}}(\Energy)\right]^{-1}\\
   &\:\times\mathbf{A}(0)e^{-i(\Energy-\hbar\Delta)t/\hbar}\dif\Energy.
\end{aligned}
\end{equation}\\

The integral can be further simplified using the residue theorem, for which is important to know the singularities of the integrand. They include poles, i.e. the solutions of the equation
\begin{equation}\label{eq:polarDet}
    \det \left[(\Energy-\hbar\Delta)\mathbf{I}+i\hbar\tilde{\mathbf{G}}(\Energy)\right]=0,
\end{equation}
as well as square-root branch points at the band edges $E_{A0}$, $E_{A1}$, $E_{B1}$, $E_{B2}$, etc. introduced in the next section.

\section{Lattice functions on the complex energy plane}\label{sec:complexEplane}

Using the Laplace transform to solve the system dynamics (see Eqn.~\eqref{eq:inverseLaplace}) makes it necessary to evaluate different lattice-related functions in the complex energy plane. This is challenging since some of these functions, such as the lattice momentum $q(\Energy)$, are multivalued and some others, such as the Bloch waves $\psi_{q(\Energy)}(0)$, are physically defined only up to a phase factor. In this section we develop an efficient way for doing so via an infinite-product representation.\\

We start by analyzing the band structure of the matter-wave vacuum using a standard textbook formula \cite{Ash&Mer},
\begin{equation}\label{eq:transmission}
\cos\frac{\pi q(\Energy)}{k}=\frac{\cos\left(\pi \sqrt{\frac{\Energy-V_b/2}{E_r}}+\text{arg}\: t(\Energy)\right)}{\lvert t(\Energy)\rvert}\equiv \Transfer(E),
\end{equation}
relating the lattice momentum $q(\Energy)$ to the transmission coefficient $t(\Energy)$ for a plane wave of energy $\Energy$ going through the isolated potential barrier described by $V(z)=\mathbf{V}_b(z)$ for $z\in [0,\pi/k]$ and 0 otherwise. Importantly, we will see that $t(E)$ does not need to be evaluated explicitly, but instead just some special points of $\Transfer(E)$ will be necessary for the analysis, as they fully determine the physical properties of the system.\\

\begin{figure}[ht]
	\begin{center}
	\includegraphics[width=1\columnwidth]{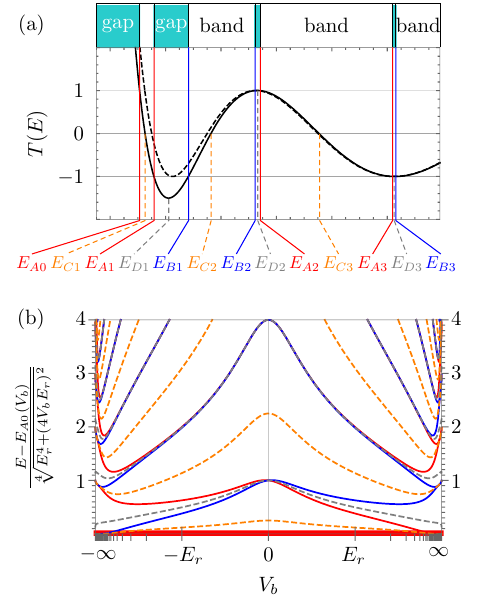}
	\caption{Characteristic energies of the matter-wave vacuum. (a) The function $\Transfer(E)$ (solid black line; example with $V_b=2.5 E_r$) determines band edges ($E_{An}$,$E_{Bn}$), the zeroes $E_{Cn}$ and the extrema $E_{Dn}$. The corresponding asymptotic expression, $\cos[\pi\sqrt{({\Energy-V_b/2})/{E_r}}]$, is shown as a black dashed line. (b) Diagram of these characteristic energies (with the same color code as in (a)) above the ground energy $E_{A0}(V_b)$ in units of an energy scale $\sqrt[4]{E_r^4+(4V_b E_r)^2}$ chosen to match the recoil energy when there is no lattice and the harmonic approximation energy split when the lattice is very deep. The abscissa $V_b$ is presented in arctangent scale.}
	\label{fig:ImportantEnergies}
	\end{center}
\end{figure}

These special points, shown in Fig.~\ref{fig:ImportantEnergies}(a), are the energies where $T(E)$ takes zero, unity, or extreme values. In particular, the energies with $\lvert \Transfer \rvert=1$ correspond to the band edges of the dispersion relation, which host \Bl\ waves carrying an integer multiple of the recoil momentum and definite parity. Correspondingly, we label the band edges with even \Bl\ waves (\Mt\ cosines) as $\lbrace E_{An}\rbrace_{n=0}^\infty$ and those with odd waves (\Mt\ sines) as $\lbrace E_{Bn}\rbrace_{n=1}^\infty$~\cite{Frenkel2001}. Finally, we label the zeroes $\lbrace E_{Cn}\rbrace_{n=1}^\infty$ and the extrema  $\lbrace E_{Dn}\rbrace_{n=1}^\infty$.\\

The ordering of these energies for different potential depths, shown in Fig.~\ref{fig:ImportantEnergies}(b), is easily understood for free-space motion $V_b=0$ and the limit $V_b\to\pm \infty$, as the lattice spectrum becomes that of a quantum harmonic oscillator. We note that flipping  $V_b\mapsto -V_b$ (as experimentally done in \cite{Stewart2020}, see Fig.~\ref{fig:TheoryVsExperiment}(c)) leaves the band structure unchanged, but swaps the parity of the edges belonging to every other energy gap. Physically, this transformation is equivalent to displacing the emitter by half a lattice period.\\

With these definitions, we can efficiently perform an analytical extension of several functions into the complex energy plane by using infinite products. For instance, consider $\Transfer(\Energy)$. Since $t(\Energy)\to 1$ as $\lvert \Energy \rvert\to \infty$, the asymptotic expression for this function readily follows from Eqn.~\eqref{eq:transmission}, and already resembles $\Transfer(\Energy)$ quite well (see Fig.~\ref{fig:ImportantEnergies}(a)). A perfect match can be achieved if one ``corrects" the zeroes of the approximation by first dividing through them  and then multiplying with the actual zeroes of T(E), which gives
\begin{equation}\label{eq:EuLanP}
\Transfer(\Energy)=\cos\left(\pi\sqrt{\frac{\Energy-V_b/2}{E_r}}\right)
\prod_{n=1}^\infty
\frac{\Energy-E_{Cn}}{\Energy-\frac{V_b}{2}-\left(n-\frac{1}{2}\right)^2E_r}.
\end{equation}
This identity is in fact guaranteed by Liouville's theorem of complex analysis \cite{LiouvilleTheorem} (assuming that $\Transfer(\Energy)$ has no complex singularities), since the quotient between the two sides of the equation is, by construction, an entire function that tends to 1.\\

Equation (\ref{eq:EuLanP}) in return allows for the analytic extension both of the lattice momentum, via $q(\Energy)=\frac{k}{\pi}\arccos \Transfer(\Energy)$, and of the energy bands $\DispersionRelationNoIndices_n(q)=\Transfer^{-1}_n \left(\cos \frac{\pi q}{k}\right)$. In particular, we note that $q(E_{Dn})$ are the branch points where the $n$ and $n+1$ bands cross.\\

An analogous reasoning can be applied for analytical extensions of other lattice functions such as $\rho(\Energy)$, $\psi_{q(\Energy)}(0)$ and $\psi'_{q(\Energy)}(0)$. Since at extreme energies ($\lvert\Energy\rvert\to\infty$) an atom travelling through the lattice potential $\mathbf{V}_b(z)$ behaves like a free particle subject to the average constant potential $V_b/2$, it is straightforward to find their asymptotic expressions, which can again be corrected further via infinite products. In the case of the density of states \hyperlink{(DoS)}{(DoS)}, this results in
\begin{equation}\label{eq:EuLanrho}
\rho(\Energy)= k\sqrt{\frac{1}{E_r(\Energy-E_{A0})}
\prod_{n=1}^\infty 
\frac{(\Energy-E_{Dn})^2}{(\Energy-E_{An})(\Energy-E_{Bn})}}.
\end{equation}

For the Bloch waves $\psi_{q(\Energy)}(0)$ taken at the emitter position, the asymptotic value for high energy is $\sqrt{k/\pi}$, and the value has to vanish at the odd band edges $E_{Bn}$ due to parity. Furthermore, Eqns.~(\ref{eq:defGtilde},\ref{eq:tightGtilde}) (or equivalently Eqn.~\eqref{eq:integralBloch}) imply that $\rho(\Energy)\psi^2_{q(\Energy)}(0)=\frac{2k}{i\pi}\sum_q {\lvert \psi_q(0)\rvert^2}/(\DispersionRelation-\Energy)$ and, given that the RHS of this expression is analytic for all $\Energy$ outside the bands, the singularities of $\psi^2_{q(\Energy)}(0)$ in this region have to be simple poles matching the zeroes $E_{Dn}$ of the DoS. This leads to the expression
\begin{equation}\label{eq:EuLanPsi}
\psi_{q(\Energy)}(0)=\sqrt{\frac{k}{\pi}
\prod_{n=1}^\infty 
\frac{(\Energy-E_{Bn})}{(\Energy-E_{Dn})}}
\end{equation}
and similarly
\begin{equation}\label{eq:EuLanPsiPrime}
\psi'_{q(\Energy)}(0)=ik^{3/2}\sqrt{\frac{\Energy-E_{A0}}{\pi E_r}
\prod_{n=1}^\infty
\frac{(\Energy-E_{An})}{(\Energy-E_{Dn})}}.
\end{equation}

Finally, we note that Eqns.~(\ref{eq:EuLanPsi},\ref{eq:EuLanPsiPrime}) can be used as initial conditions in the \Mt\ equation to obtain the value of the Bloch wave at any other point through  $\psi_{q(\Energy)}(z)=\psi_{q(\Energy)}(0)C(\Energy,z)+i\psi'_{q(\Energy)}(0)S(\Energy,z)$ where $C$ ($S$) is an entire function in both of their arguments, (anti-)symmetric in $z$ corresponding to the unnormalized Mathieu cosine (sine) function.\\


\section{Single-emitter decay ($N=1$)}\label{sec:N=1}

The decay of an isolated emitter for the case of coupling to a single, sinusoidal band or a parabolic band edge has been analyzed in earlier theoretical works \cite{Stewart2017, Lombardo2014, Calajo2016}. Using our formalism, we now generalize this treatment to include infinitely many (non-)sinusoidal bands, arising for arbitrary lattice depth $V_b$. For simplicity, we consider the case $V_b>0$ and disregard band gaps beyond an arbitrary cutoff index $\Lambda$ (in practice $\Lambda$ can always be chosen to be small, given that the band gaps quickly get narrower).\\

As seen in the previous section, the dynamics $A(t)$ of a tightly confining emitter is governed by the bath correlation function $ \tilde G_{j\text{-}J}(E)\equiv \tilde G_{0}(E)$ obtained from Eqns.~\eqref{eq:tightGtilde}, \eqref{eq:EuLanrho} and \eqref{eq:EuLanPsi}. To evaluate $\tilde G_{0}(E)$, we first introduce the (truncated) products $\Pi_A(\Energy)=\prod_{n=0}^\Lambda (\Energy-E_{An})$ and $\Pi_B(\Energy)=\prod_{n=1}^\Lambda (\Energy-E_{Bn})$, as well as their ratio $\Pi_{B/A}(\Energy)=\Pi_B(\Energy)/\Pi_A(\Energy)$. Using these products, the bath correlation function can then be approximated as $\hbar\tilde G_{0}(E)\approx\kappa\sqrt{\Pi_{B/A}(\Energy)}$, with the coupling constant
$\kappa=(\Omega/2)^2 \hbar a_{ho} \sqrt{2\pi m}$. To avoid ambiguity, we consider that all square roots $\sqrt{\cdots}$ in this section give back complex numbers with argument in $(-\pi/2,+\pi/2]$.\\

\begin{figure}[t]
	\begin{center}
	\includegraphics[width=1\columnwidth]{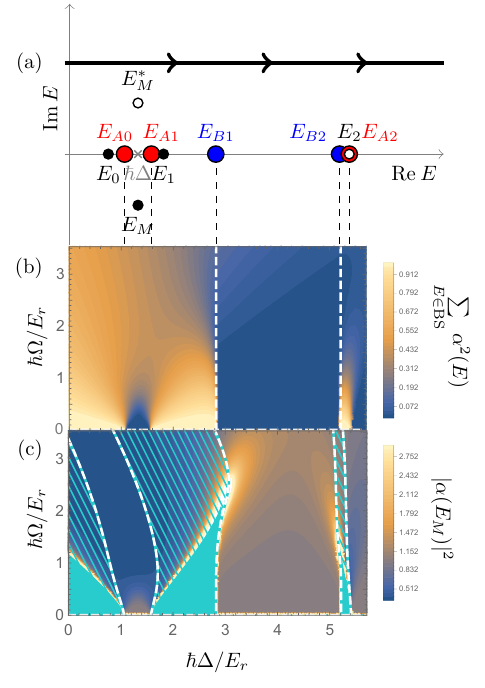}
	\caption{Spectral emission properties of a tight emitter. (a) Example of poles and branch cuts (gray dashed lines) of the inverse Laplace transform~(\ref{eq:inverseLaplace}) for $V_a=20E_r$, $V_b=2.5E_r$, coupling $\hbar\Omega=E_r$ and a detuning $\Delta$ (in gray) resonant with the first energy band. 
	Black dots correspond to physical poles, whereas the white dots are unphysical and do not contribute to the dynamics. The red and blue branch points are the band edges hosting even and odd Bloch waves, respectively. The thick black line represents the integration contour of the transform.
	(b) The sum of the squared residues of the bound states --an indicator of the red population that reminds bounded after emission-- is presented as a color-map on the $\Delta$-$\Omega$ plane. Regions separated by the white dashed vertical lines $\lbrace\hbar\Delta=E_{Bn}\rbrace_{n= 1}^\infty$ have a different number of bound states. (c) Squared norm of the residue corresponding with the Markovian pole $\Energy_M$ on the $\Delta$-$\Omega$ plane. In the solid cyan regions there is no Markovian pole, whereas in the cyan dashed regions the pole is in the lower sheet.}
	\label{fig:BSphaseDiagram}
	\end{center}
\end{figure}

To find the poles of the inverse Laplace transform ~(\ref{eq:inverseLaplace}), we multiply Eqn.~\eqref{eq:polarDet} with its algebraic conjugate; the poles then correspond to zeroes of the polynomial $(\Energy-\hbar\Delta)^2\Pi_A(\Energy)+\kappa^2\Pi_B(\Energy)$. By considering the degree of this polynomial and its changes in sign between band edges, it is easy to see that
there are no bound states in the continuum (BIC) \cite{Hsu2016}, whereas each band gap (including $E<E_{A0}$) contains at least one pole. This differs from the multi-emitter case $1<N<\infty$, where subradiance and retardation effects can lead to a BIC \cite{Sinha2020, Calajo2019}. For sufficiently weak couplings ($\kappa\ll E_r^{3/2}$), these in-gap poles can be approximated as $\Energy_n\approx E_{An}-\kappa^2 \Pi_{B}(E_{An})\:/\:[(E_{An}-\hbar\Delta)^2\prod_{m\neq n}(E_{An}-E_{Am})]$. 
There are two additional poles, approximately $\hbar\Delta\mp i\kappa \sqrt{\Pi_{B/A}(\hbar\Delta)}$, which lie in the same gap if they are real; otherwise one of them ($E_M$) has negative imaginary part and can lead to Markovian (exponential) decay of the emitter, and the other ($E_M^*$) is its complex conjugate.\\

In determining the spectral decay properties, see Fig.~\ref{fig:BSphaseDiagram}, we see that not all of these poles contribute towards the residue theorem: due to the square root singularities at the band edges, we can visualize the integrand domain as a Riemann surface consisting of an `upper sheet' where the integration paths in the complex plane are located and a `lower sheet' on the other side of the branch cuts. Only the poles on the upper sheet will contribute towards the residue theorem and have a physical interpretation.\\

For positive lattice depths $V_b>0$ and Markovian couplings $\Omega\ll\min_{q\in\mathbb{R}}\vert\Delta-\DispersionRelation/\hbar\rvert$, the pole $\Energy_n$ is in the upper sheet only when $\operatorname{sign}\lbrace\hbar\Delta-E_{An}\rbrace=(-1)^n$. Of the two extra poles, only one is in the upper sheet; in particular $E_M^*$ cannot be there because its positive imaginary part would lead to exponential growth of the population (see Eqn.~\eqref{eq:Aevolution}).\\

For larger couplings, there is still a change of sheets for one of the poles living in the $n^{th}>0$ gap as $\hbar\Delta$ crosses the value $E_{Bn}$. On the other hand, by increasing the coupling it is possible to make two lower poles co-located in a gap combine into a double pole and then split into a Markovian pole $\Energy_{M}$ and its conjugate $\Energy_{M}^*$. Whereas $\Energy_{M}^*$ remains always unphysical, $\Energy_{M}$ can make it to the upper sheet, as depicted in Fig.~\ref{fig:BSphaseDiagram}(c).\\

This figure reveals behavioral differences between the decay next to a band edge hosting even Bloch waves and one hosting odd ones. Whereas the former behaves as expected from previous studies \cite{Stewart2017}, the latter displays an increase in the Markovian component of the decay at non-Markovian couplings. One would naively expect that reabsorption and emission scale equally with the vacuum coupling; however this is not the case here as BS formation is suppressed for these parameters (see Fig.~\ref{fig:BSphaseDiagram}(b)). Despite the phenomenological similarities, this {\it ultra-Markovian} emission is not to be confused with superradiance, as a single emitter is enough to create this effect. For illustration purposes, let us consider the following example.  Under the same conditions as Fig.~\ref{fig:BSphaseDiagram}(b,c), a quantum emitter with detuning $\hbar\Delta=E_{B1}+\tfrac{1}{5}E_r$ and coupling $\hbar\Omega=\tfrac{5}{2}E_r$ emits half of its population at only $t=0.52\hbar/E_r$ and it emits more than $90\%$ ultimately. In contrast, while the initial decay is very similar at $\hbar\Delta=E_{A0}+\tfrac{1}{5}E_r$ and $\hbar\Omega=\tfrac{5}{2}E_r$, most of the radiation is reabsorbed at $t=2.5\hbar/E_r$ and the red population subsequently oscillates with large amplitude as the system cycles between emission and reabsorption.\\

The integration path of Eqn.~\eqref{eq:inverseLaplace} can be adapted to the singularities described in Fig.~\ref{fig:BSphaseDiagram}(a) by circling around the physical poles and branch cuts of the integrand \cite{Note2}, leading to the time evolution 
\begin{equation}
\begin{aligned}\label{eq:Aevolution}
A(t)=&\sum_{\substack{\Energy\in\text{upper}\\ \text{poles}} }\EmitterResidue(\Energy)e^{i(\Delta-\Energy/\hbar)t}
\\
&-\frac{i\kappa}{\pi}\sum_{n=0}^\Lambda (-1)^n  e^{i(\Delta-E_{An}/\hbar)t} \Integral \left(E_{An},t\right)
\\
&+\frac{i\kappa}{\pi}\sum_{n=1}^\Lambda (-1)^n  e^{i(\Delta-E_{Bn}/\hbar)t}\Integral \left(E_{Bn},t\right)
\end{aligned}
\end{equation}
where
\begin{equation}
    \EmitterResidue(\Energy)= \frac{2(\Energy-\hbar\Delta)}{2(\Energy-\hbar\Delta)+\kappa^2\frac{\dif}{\dif\Energy}\Pi_{B/A}(\Energy)}
\end{equation}
denotes the residue of the poles and the branch contribution
\begin{equation}
    \Integral \left(\Energy,t\right)= \int_{0}^\infty \frac{
\exp\left({-{\zeta t}/{\hbar}}\right)
\sqrt{\Pi_{B/A}(\Energy-i\zeta)}}
{(\Energy -i\zeta -\hbar\Delta)^2+\kappa^2\Pi_{B/A}(\Energy-i\zeta)}\dif\zeta
\end{equation}
is well defined unless $\Energy_M$ stands on the branch cut. The branch contribution $\Integral \left(\Energy,t\right)$ tends to zero as the time $t$ increases but in a non-exponential fashion, making the decay non-Markovian. The only contributions persistent in time, i.e. the bound states (BS), are the real upper poles residing in the band gaps.\\

\begin{figure*}[]
	\begin{center}
	\includegraphics[width=1\textwidth]{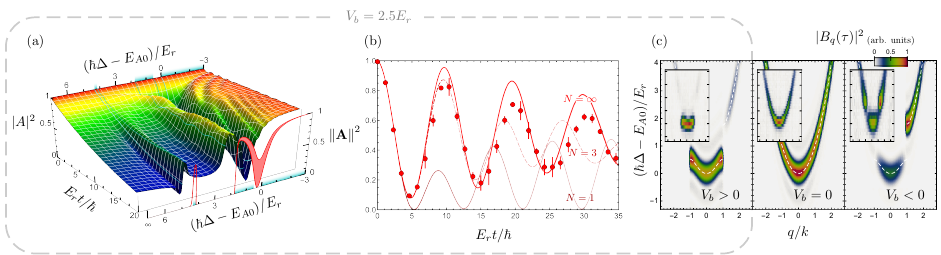}
	\caption{Comparison between theory and the experiments \cite{Stewart2020,Krinner2018}. (a) Detuning dependence of the population decay $\lvert A(t) \rvert^2$, from one tightly-confining emitter to a sinusoidal lattice of potential depth $V_b=2.5E_r$ and through a coupling $\kappa=0.082 E_r^{3/2}$. Cyan marks indicate the position of the energy gaps. The time slice $t=\infty$ represents (in red) the range of asymptotic population that remains bound after decay. (b) Decay curves as in (a) but for a set detuning in the center of the first energy band, $\hbar \Delta=1.32 E_r$ and for an array of $N=1,3$ and $\infty$ emitters, with only the central emitter originally excited. The red dots indicate the experimental values measured for the same parameters \cite{Stewart2020}. (c) Quasi-momentum distribution $\lvert B_q(\tau)\rvert^2$ predicted for the emission by a single tightly-confining emitter with $\kappa E_r^{-3/2}=0.015$, 0.032 and 0.015 into potentials with depth $V_{b}/E_r=2.5$, 0 and $-2.6$ (respectively) at a fixed time $E_r\tau/\hbar=9.24$. The insets reproduce the experimentally observed distributions for $V_b=0$ \cite{Krinner2018} and $V_b\neq0$ \cite{Stewart2020} measured for the same parameters.}
	\label{fig:TheoryVsExperiment}
	\end{center}
\end{figure*}

The general features of the resulting time evolution are shown in  Fig.~\ref{fig:TheoryVsExperiment}(a). It is mostly Markovian for detunings deep inside the bands and non-Markovian around the band edges; there is no decay deep inside the gaps. We note that that emission is suppressed at the band edges $E_{Bn}$ of odd parity with respect with the edges $E_{An}$, whose Bloch waves have the same parity as the emitter.\\

After the emission, the emitter population can oscillate by the beating of various bound states. This effect, which has also been measured experimentally \cite{Stewart2020}, is most noticeable in the center of the first energy band. We compare our calculation with the experimental data in Fig.~\ref{fig:TheoryVsExperiment}(b).\\

Although the one-emitter model ($N=1$) matches the observed decay dynamics at short times, it underestimates the amount of subsequent reabsorption seen in the experiment, in which the optical lattice provided an array of emitters. As already discussed in \cite{Stewart2020}, the subsequent oscillations are dominated by reabsorption as the  emitted radiation spreads across the emitter array. Using the formalism developed in this paper, the presence of neighboring ground-state emitters (i.e. empty lattice sites) surrounding an excited emitter can readily be taken into account in Eqn.~\eqref{eq:inverseLaplace} via an analogous approach, and already working with a lattice of three sites ($N=3$) shows a marked improvement for the second oscillation. While it generally gets harder to analyze and numerically solve larger arrays, a special case is $N=\infty$, analyzed in Sec.~\ref{sec:polaritons}. As seen in the figure~\ref{fig:TheoryVsExperiment}(b), the overall agreement with the experiment qualitatively improves further to longer time scales, but residual deviations persist. They are likely due to differences in the initial state (the experiment worked with a sparsely populated array with more than one excitation) and collisions between atoms.\\

An analogous formula to \eqref{eq:Aevolution} can be written for the time evolution $B_q(t)$ of the emitted modes,
\begin{equation}
B_q(t)=-\frac{\hbar\Omega\gamma_q}{4\pi i}\int_{-\infty+i0^+}^{+\infty+i0^+}\frac{e^{-i(\Energy-\DispersionRelation)t/\hbar}\ \dif\Energy/\left(\Energy-\DispersionRelation\right)}{\Energy-\hbar\Delta+i\kappa\sqrt{\Pi_{B/A}(\Energy)}}.
\end{equation}
The main difference is that the integration displays an additional pole at energy $\DispersionRelation$. Importantly, this real pole does not correspond to a bound state (discussed further in the next section), but to a mode that has completely abandoned the emitter and keeps travelling free indefinitely. This causes similarity between the emission spectrum and the dispersion relation of the medium (see Fig.~\ref{fig:TheoryVsExperiment}(c)). We note that at smaller couplings, the applicability of the $N=1$ theory extends to longer times, which was the case for the parameters of Fig.~\ref{fig:TheoryVsExperiment}(c). More generally, a multi-emitter analysis of the emitted modes is accessible through Eqn.~\eqref{eq:EOMsystem}.\\

\section{Bound states}\label{sec:BoundStates}

As seen in the previous section, the long-time dynamics of the quantum emitter is dictated by the presence of bound states. BS have been broadly studied in the literature \cite{Shi2016} and in 1D are limited to be short ranged \cite{SBurillo2020}, which is the reason why they are often depicted as decaying exponentially in space \cite{Stewart2017, Shi2018, Shi2016, deVega2008, Calajo2016} while the detailed features are often overlooked. However, the reader might expect that the presence of a lattice potential for the radiated modes induces a corrugation of this evanescent wave \cite{Bykov1975}. Even more strikingly, recent experimental work \cite{Stewart2020} found that time-of-flight distributions of these BS can possess two sharp peaks at opposite momenta, suggesting common features with stationary waves despite the absence of boundaries. In this section we clarify these apparent inconsistencies by computing the exactly spatial distribution of the BS and presenting a simple physical picture that encompasses all these effects.\\

The BS wavefunction $\ket\psi(t)$ is defined in the interaction picture (see Eqn.~\eqref{eq:psi}) by 
\begin{equation}\label{eq:BSwavefunction}
\left\lbrace\begin{array}{l}
  A(t)=A(0)e^{i(\Delta-\Energy_{BS}/\hbar)t}   \\
B_q(t)=\frac{\hbar\Omega}{2(\Energy_{BS}-\DispersionRelation)}\gamma_q A(0)e^{i(\DispersionRelation-\Energy_{BS} )t/\hbar},
\end{array}\right.
\end{equation}
where $A(0)=\left[1+\sum_q\lvert\gamma_q\rvert^2\left(\frac{\hbar\Omega/2}{\Energy_{BS}-\DispersionRelation}\right)^2\right]^{-1/2}$ is the normalization constant. The red part of this expression follows from isolating the contribution of a single pole in Eqn.~\eqref{eq:Aevolution}, whereas the blue part and the equation~\eqref{eq:polarDet} for the BS energy $\Energy_{BS}$ follow from integrating~(\ref{eq:EOMsystem}).\\

The spatial distribution of this bound state is better understood in the \Schr\ picture $\ket{\psi(t)}_S=e^{-i\hat{H}_0 t/\hbar}\ket{\psi(t)}$, where position is a time-independent operator. The blue part of $\ket{\psi(t)}_S$, has a position-space wave function $B_S(z,t)=\sum_q B_{S,q}(t)\psi_q(z)$ which can be integrated in a very similar way to the derivation of~(\ref{eq:modalGtilde}), leading to
\begin{widetext}
\begin{equation}\label{eq:spatialBS}
\begin{aligned}
B_S&(z,t)=-i\rho(\Energy_{BS})\psi_{q(\Energy_{BS})}^2(0)\frac{\sqrt{2a_{ho}}\hbar\Omega A(0)\pi^{5/4}}{8k}e^{-i(\omega_b+\Energy_{BS}/\hbar )t}\sum_{m,r=-\infty}^{+\infty}\upsilon_m(\Energy_{BS})\upsilon_r(\Energy_{BS})
\\
&\times\: e^{-\frac{(q(\Energy_{BS})+2kr)^2a_{ho}^2}{2}}\left[e^{-i(q(\Energy_{BS})+2km)z}\operatorname{erfc}\left(-i\frac{(q(\Energy_{BS})+2kr)a_{ho}}{\sqrt{2}}+\frac{z}{\sqrt{2}a_{ho}}\right)+(z\leftrightarrow -z)\right].
\end{aligned}
\end{equation}
\end{widetext}

\begin{figure}[b]
	\begin{center}
	\includegraphics[width=1\columnwidth]{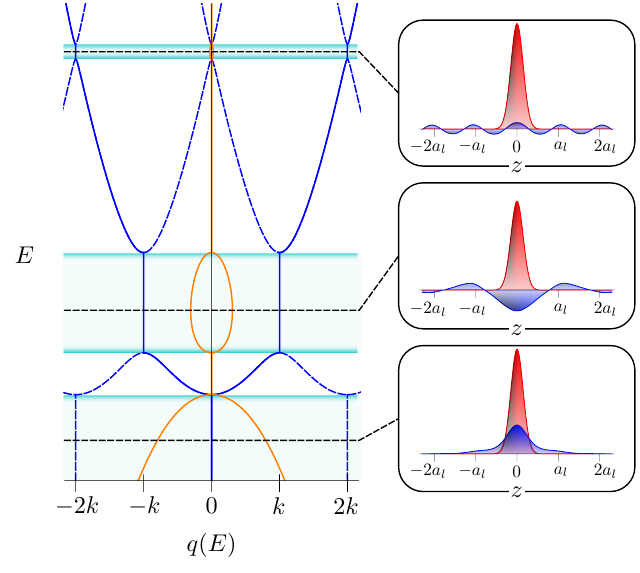}
	\caption{Spatial shape of the bound states for the same parameter values as Fig.~\ref{fig:TheoryVsExperiment}(a). On the left, the real (blue) and imaginary (orange) parts of the analytic extension of $q(E)$ are presented for various energies in the bands (white) and gaps (cyan). The solid lines denote the momenta that contribute the most in the composition of the bound state. The horizontal black dashed lines denote three different possible BS energies at different gaps. They intersect with the lattice momenta that conform the evanescent waves around the emitter, i.e. the bound states, which are shown on the right in spatial units given by the lattice constant $a_l=\pi/k$.\vspace{-0.75cm}}
	\label{fig:boundStates}
	\end{center}
\end{figure}

This expression might seem cumbersome, but we note that it is real at $t=0$ and symmetric; that the prefactor is regular also when $\Energy_{BS}=E_{Dn}$, due to the cancellation of the zero of $\rho(\Energy_{BS})$ with the pole of $\psi_{q(\Energy_{BS})}^2(0)$; that the time evolution is that of an eigenstate of the system (even though the Hamiltonian is not time-independent due to the external coupling, which causes the red part of the BS to have a frequency lower than the blue by $\omega_\mu$); and that the dominant modes of the sums are $\upsilon_0(\Energy_{BS})$ and $\upsilon_{-n}(\Energy_{BS})=\upsilon_0^*(\Energy_{BS})$ (with $n$ numbering the gap of $\Energy_{BS}$) when the blue lattice is shallow as they correspond with the momenta that are closest to the free-particle dispersion relation (see Fig.~\ref{fig:boundStates}). The BS is localized at the center of the site and decays to both sides with an inverse exponential decay length $\operatorname{Im} q(\Energy_{BS})$ and a wavenumber $\operatorname{Re} q(\Energy_{BS})=nk$, and it is consistent with those presented in Refs.~\cite{Stewart2017, Bykov1975}. In summary, expression \eqref{eq:spatialBS} shows that the BS is a linear combination of evanescent waves whose momenta match the analytic extension to the band gaps of the blue lattice dispersion relation, as presented in Fig.~\ref{fig:boundStates}.\\

Looking back at the spectrum of the system described in the previous section, these bound states may be created adiabatically by choosing a detuning $\Delta$ resonant with the $n^{th}$ band gap and increasing the  vacuum coupling strength $\Omega$ adiabatically, as empirically shown in Refs.~\cite{Krinner2018,Stewart2020}. The adiabatic condition consists of the change in these parameters being slower than the time scale defined by the energy difference between the instantaneous BS energy and the nearest energy edge. This imposes two restrictions for adiabatic excursions in the $\Delta$-$\Omega$ plane (see Fig.~\ref{fig:BSphaseDiagram}(b)): $\Omega$ must not be brought back to 0 while $\Delta$ is in some region other than the $n^{th}$ band gap since the BS energy would cross $E_{An}$, and $\Delta$ must not cross the corresponding odd band edge $E_{Bn}/\hbar$ since the BS energy would also cross $E_{Bn}$.\\

The procedure of adiabatic creation allows for testing experimentally the BS probability distributions derived in this section. The lattice-momentum distribution $\lvert B_q (t)\rvert^2=\lvert B_{S,q} (t)\rvert^2$ of Eqn.~\eqref{eq:BSwavefunction} matches with the band-map measurements of Refs.~\cite{Krinner2018} (for the case $V_b=0$; see also \cite{Stewart2017}) and Ref.~\cite{Stewart2020} (case $V_b=2.5 E_r>0$). Furthermore, signatures of the real-space distribution $\lvert B_S (z,t)\rvert^2$ of Eqn.~\eqref{eq:spatialBS} might be accessed by highly resolving \textit{in situ} imaging techniques \cite{Bakr2009}.\\

\section{Matter-wave polaritons}
\label{sec:polaritons}

In general, as the number of emitters in the array increases, the spectrum of the system becomes more elaborate, but the case of infinitely many emitters is an exception to this. As $N\to\infty$, the Hamiltonian regains the lattice periodicity and states with different lattice momentum $q$
decouple via Bloch's theorem. In our formalism this is reflected in the fact that vectors of the form $\mathbf{A}(0)=(\ldots,e^{iq\pi(j-1)/k},e^{iq\pi j/k},e^{iq\pi(j+1)/k}\ldots)^{\operatorname {T} }$ are eigenstates of $\tilde{\mathbf{G}}(\Energy)$ while being independent of $\Energy$. We will refer to these states as {\it matter-wave polaritons} \cite{Kwon2021}, in analogy to of their quantum optical counterparts \cite{Carusotto2013, Shi2018, Basov2021}. While photons are also involved in this new type of quasiparticle by constituting the optical lattice, the usual role of a photon in a polariton is taken over by matter waves. In the regime $\lvert V_b \rvert\ll E_r\ll\lvert V_a \rvert$, the dispersion relation of the radiated modes is nearly quadratic and matter-wave polaritons are similar to exciton polaritons \cite{Carusotto2013,Schneider2017}, whereas for $E_r\ll\lvert V_b \rvert\ll\lvert V_a \rvert$ the now strongly confining optical lattice for blue atoms forms the analogue of a  coupled cavity array in circuit QED \cite{Hartmann2006, Greentree2006,Noh2016, Hartmann2016,Fitzpatrick2017,Ma2019}.\\

Taking the Fourier transform $\FA=\sum_j A_j e^{-i\pi j q/k}$ as a change to the decoupled basis, Eqn.~\eqref{eq:inverseLaplace} simplifies to
\begin{equation}\label{eq:PolaritonInvLaplace}
    \begin{aligned}
    \FA(t)=\frac{-1}{2\pi i}\int_{-\infty+i\epsilon}^{+\infty+i\epsilon}
    \frac{\FA(0)e^{-i(\Energy-\hbar\Delta)t/\hbar}}
    {\Energy-\hbar\Delta+i\hbar\tilde{g}_q(\Energy)}\dif\Energy,
\end{aligned}
\end{equation}
where the eigenvalue
\begin{equation}\label{eq:PolaritonEnergies}
    i\hbar\tilde{g}_q(\Energy)=\left(\frac{\hbar\Omega}{2}\right)^2\sum_{n=-\infty}^{+\infty}\frac{\lvert \gamma_{q+2kn}\rvert^2}{\DispersionRelationNoIndices_{q+2kn}-\Energy}
\end{equation}
follows immediately by applying the definition of eigenvalue to~(\ref{eq:defGtilde}) and using the identity $\sum_{j=-\infty}^{+\infty} e^{i(q'-q)\pi j/k}=2k\sum_{n=-\infty}^{+\infty}\delta(q'-(q+2kn))$. The integrand has thus become a meromorphic real function, free of the branch cuts at the band edges and free of complex poles. This indicates that radiation cannot escape the emitter array, which is to be expected since the array is infinite.\\

\begin{figure}[t]
	\begin{center}
	\includegraphics[width=1\columnwidth]{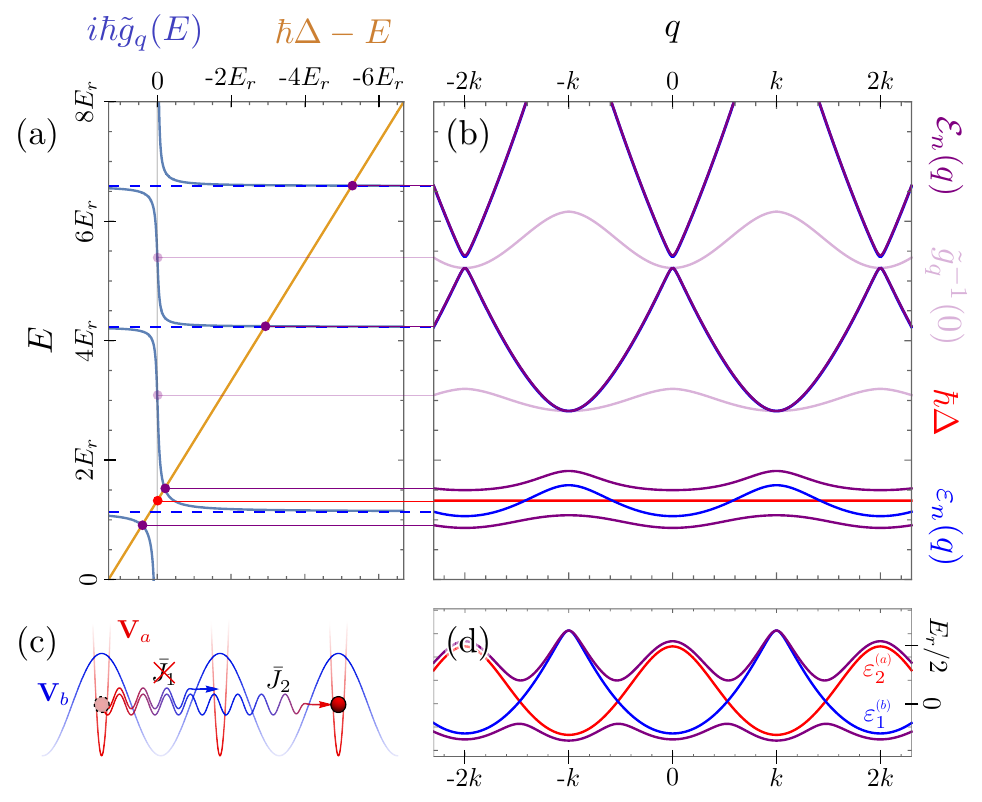}
	\caption{Band structure of the polaritons. (a) Graphical solution of Eqn.~\eqref{eq:PolaritonEnergies} for the polariton bands at a particular quasimomentum, $q_0=-2.3k$, and with the same parameter values as Fig.~\ref{fig:TheoryVsExperiment}(b). Polariton energies are located at the intersection between the orange and blue lines. (b) Resulting polariton band structure $\mathcal{E}_n(q)$ (in purple), for the band structure $\DispersionRelationNoIndices_n(q)$ (in blue) and detuning $\hbar\Delta$ (in red) specified in (a). (c) Schematic depiction of a particle in a state-dependent lattice with $V_a=12 E_r$ and $V_b=-0.4949E_r$ hopping two lattice sites due to the coupling $\hbar\Omega=0.626 E_r$ between states. (d) Energy bands corresponding to the situation depicted in (c), with a detuning of $0.0742 E_r$ between the centers of the ground band $\DispersionRelationNoIndices_1^{\scalebox{0.5}{(b)}}(q)$ and the excited band $\DispersionRelationNoIndices_2^{\scalebox{0.5}{(a)}}(q)$. The (approximate) doubling in periodicity of the resulting polariton bands (in purple) is an indicator of the double hopping.}
	\label{fig:PolaritonBands}
	\end{center}
\end{figure}

Since all of the poles are real, we follow a suggestion in ~\cite{Bykov1975} and visualize Equation \eqref{eq:PolaritonEnergies} as shown in  Fig.~\ref{fig:PolaritonBands}(a), in order to locate all of the solutions $\mathcal{E}_n(q)$ that physically correspond to the polariton energy bands. Some simple properties that follow from this graph are that polariton bands neither cross each other nor the original energy bands ($\mathcal{E}_1(q)<\hbar\Delta$, $\DispersionRelationNoIndices_{n-1}(q)<\mathcal{E}_n(q)<\DispersionRelationNoIndices_n(q)$ for $n\geq2$), although they might cross the detuning at the points $\tilde{g}_q^{-1}(0)$ where the couplings to different bands cancel mutually. Excited polariton bands tend to these points in the limit of very large coupling; otherwise they soon approximate the energy bands ($\mathcal{E}_{n\gg1}(q)\approx \DispersionRelationNoIndices_{n-1}(q)$).\\

The resulting band structure (purple lines in Fig.~\ref{fig:PolaritonBands}(b)) is exotic and cannot be obtained by a simple periodic potential. An indicator of this is the positive effective mass of both the ground and the first excited polariton band near $q=0$.\\

The motional properties associated with these bands follow from applying the residue theorem to \eqref{eq:PolaritonInvLaplace}, which leads to
\begin{equation}\label{eq:polaritonSoln}
    \FA(t)=\sum_{n=1}^\infty \PolaritonResidue_n(q) e^{i(\Delta-\mathcal{E}_n(q)/\hbar)t}\FA(0)
\end{equation}
where the residues are given by
\begin{equation}
    \PolaritonResidue_n(q)=(1+i\hbar \tilde{g}'_q(\mathcal{E}_n(q)))^{-1}.
\end{equation}
The residues for bands far away from the detuning are negligible, rendering the structure of higher bands/gaps irrelevant for the dynamics. Alternatively, these results may also be obtained by decoupling the Hamiltonian \cite{Shi2018} or as an ansatz in the \Seq, while taking into consideration that these residues satisfy the normalization condition ($\PolaritonResidue_n\in[0,1]$ and $\sum_{n=1}^\infty \PolaritonResidue_n(q)=1$), energy conservation ($\sum_{n=1}^\infty \PolaritonResidue_n(q)\mathcal{E}_n(q)=\hbar\Delta$) and $\sum_{n=1}^\infty \PolaritonResidue_n(q)/(\DispersionRelationNoIndices_m(q)-\mathcal{E}_n(q))=0$. With this solution of the system, we can explain the dynamical behaviour of the experimental data \cite{Krinner2018,Stewart2020} at longer evolution times $t\geq \hbar/J_b$ (see Fig.~\ref{fig:TheoryVsExperiment}(b), case $N=\infty$).\\

Moreover, as different quasi-momenta are decoupled unless they differ by an even multiple of the recoil momentum $k$, one can define a periodic momentum-dependent detuning $\Delta\equiv\Delta(q)$ to account more accurately for the dispersion relation $\DispersionRelationNoIndices_n^{\scalebox{0.5}{$(a)$}}(q)$ for the $\ket{a}$ states in the $n^{\text{th}}$ band, while also modifying the \FC\ overlap into $\gamma_q=\esc{\psi^{\scalebox{0.5}{$(b)$}}_q}{\psi^{\scalebox{0.5}{$(a)$}}_{n,q}}$. This allows more customization of the resulting polariton bands $\mathcal{E}_{n'}(q)$, whose hopping rates
\begin{equation}
\bar{J}_j^{(n')}=-\int_{-k}^{k} \frac{\dif q}{2k}\mathcal{E}_{n'}(q)e^{i\pi j q/k}
\end{equation}
can be freely tuned. An extreme example of this is shown in Fig.~\ref{fig:PolaritonBands}(c,d) where the first band of the $\ket{b}$ states is coupled with the second band of the $\ket{a}$ in a way that the polaritons hop two lattice sites at a time, without going through the intermediate site at all ($\bar J_1=0$, $\bar J_2\neq0$). 
This opens the possibility for the analog simulation of $J_1$-$J_2$ quantum spin models \cite{Chandra1988,Dagotto1989} with ultracold bosons in 1D, by inducing effective spin interactions \cite{Kuklov2003, Altman2003,Liao2021}. We note that  frustration \cite{Balents2010}, a core feature of the $J_1$-$J_2$ model, in our system has kinetic origin and is generated by coupling bands of opposite effective mass~\cite{Strater2015}.
\\

\section{Conclusions}
\label{sec:conclusion}

We have shown the importance of having multiple emitters coupled to multiple band modes in ultracold-atom realizations of waveguide QED. Already on the single emitter level, the intricate band structure enriches the spectrum, partially due to the presence of energy edges hosting odd Bloch waves (which present an unexpected enhancement of the Markovian part of the decay in the non-Markovian regime), and partially due to the presence of multiple bound states in different band gaps. Our analysis was made possible by what to our knowledge are novel, simple analytical expressions for multiple well-known functions describing the dynamical properties of a particle in a lattice potential. The fast convergence of these expressions and their applicability on the whole complex energy plane make them relevant in contexts beyond the scope of this work, although there is no trivial generalization to higher dimensional lattices. And finally we have studied polariton formation in the matter-wave context, showing that this system not only can act as an analog simulator of photonic phenomena, but also as a wider platform for studying low-dimensional frustrated systems.\\


\begin{acknowledgments} We thank M. Stewart for detailed early discussions and M. G. Cohen for a critical reading of the manuscript. This work was supported by the National Science Foundation under grant No. PHY-1912546, with additional funds from the SUNY Center for Quantum Information Science on Long Island.
\end{acknowledgments}

\begin{appendix}
\section{Integrating over lattice momenta}\label{app:planeDecomposition}

\begin{figure}[b]
\includegraphics[width=0.35\textwidth]{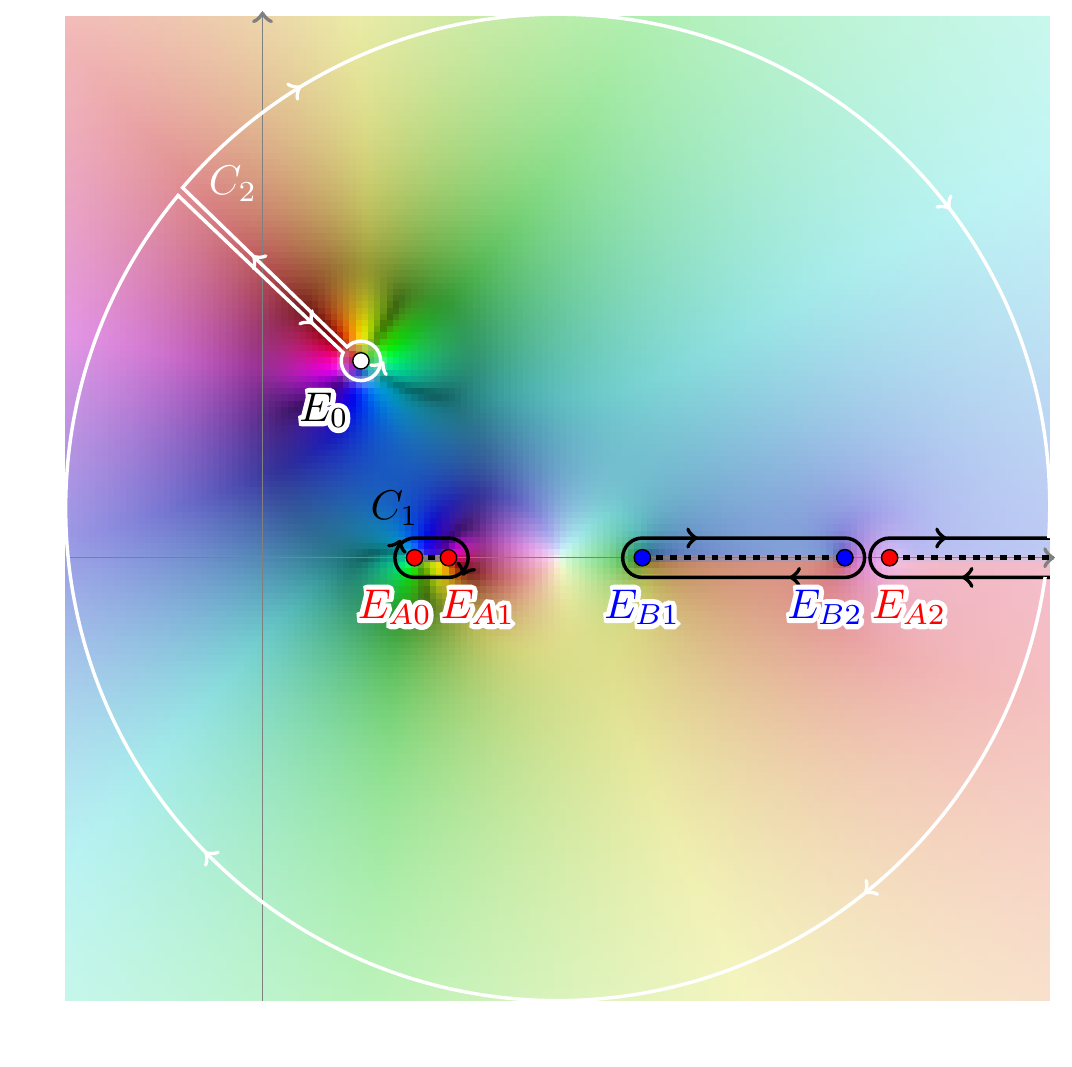}
\caption{\label{fig:integrandContours} Domain coloring plot of the integrand of the RHS of \eqref{eq:integrand} for $J=j=0$, $kz'=0.5<kz''=0.7$, $V_s=4E_r$ and $\Energy_0/E_r=1+2i$.}
\end{figure}

The study of spontaneous emission into a lattice calls for integration over the lattice momenta both in the dynamics (see Eqn.~\eqref{eq:defGtilde}) and in the spatial distribution of the bound state,
\begin{equation}
    B_S(z,t)=\sum_q B_{S,q}\psi_q(z)\propto \int_{-\infty}^{+\infty} \frac{\gamma_q\psi_q(z)}{\Energy_{BS}-\DispersionRelation}\dif q.
\end{equation}
In order to simplify them, we propose expressing the \FC~overlaps in their integral form and solving first the integral
\begin{equation}\label{eq:integrand}
\begin{aligned}
    \int_{-\infty}^{+\infty} \frac{{\psi}_{q}^*(z'){\psi}_{q}(z'')e^{iq(z_j-z_J)}}
    {\Energy_0-\DispersionRelation}\dif q
    =
    \\
    \oint_{C_1} \frac{{\psi}_{q(\Energy)}(-z'){\psi}_{q(\Energy)}(z'')\rho(\Energy)e^{iq(\Energy)(z_j-z_J)}}
    {2(\Energy_0-\Energy)}\dif\Energy
    \end{aligned}
\end{equation}
where $C_1$ are contours circling the bands (see Fig.~\ref{fig:integrandContours}). By using Sec.~\ref{sec:complexEplane}, it follows that the integral has a symmetry $(z',z_J)\leftrightarrow(z'',z_j)$ and the integrand has only a simple pole in $\Energy_0$ and bi-valued branch cuts at the band edges. Changing the integration contour to $C_2$ and noticing that the integrand is asymptotically dominated by ${\psi}_{q(\Energy)}^*(z'){\psi}_{q(\Energy)}(z'')e^{iq(\Energy)(z_j-z_J)}\sim {\exp[ik(z''+z_j-z'-z_J)\sqrt{(\Energy-V_b/2)/E_r}]}$, we find that the outermost circumference of $C_2$ vanishes in the upper sheet (Im $q(\Energy)>0$) when $z''+z_j-z'-z_J>0$ (the opposite case follows by the aforementioned symmetry), leaving only the pole contribution of $\Energy_0$:
\begin{widetext}
\begin{equation}\label{eq:integralBloch}
\begin{aligned}
    \int_{-\infty}^{+\infty} \frac{{\psi}_{q}^*(z'){\psi}_{q}(z'')e^{iq(z_j-z_J)}}
    {\Energy_0-\DispersionRelation}\dif q
    =
    -\pi i{\psi}_{q(\Energy_0)}(-z'){\psi}_{q(\Energy_0)}(z'')\rho(\Energy_0)e^{iq(\Energy_0)(z_j-z_J)}H(z''+z_j-z'-z_J)
    +\left(
    \begin{array}{c}
         z'\leftrightarrow z''\\
         z_J\leftrightarrow z_j
    \end{array}
    \right),
\end{aligned}
\end{equation}
\end{widetext}
with $H(z)=0,1/2,1$ if $z<,=,>0$ (respectively)  denoting the Heaviside step function.\\

The overlapping integral with the Gaussian emitter $\phi_0(z)$ can then be solved exactly after Fourier-decomposing the Bloch waves into plane waves (see Eqn.~\eqref{eq:upsilons}), leading to expressions (\ref{eq:modalGtilde}, \ref{eq:spatialBS}).\\
\vspace{5cm}

\end{appendix}


\bibliographystyle{apsrev}

\end{document}